\documentclass[tightenlines,aps,twocolumn,
showpacs,nofootinbib,floatfix]{revtex4}

\usepackage{float}
\newcommand{\be}{\begin{equation}}
\newcommand{\ee}{\end{equation}}
\newcommand{\ba}{\begin{eqnarray}}
\newcommand{\ea}{\end{eqnarray}}

\newcommand{\ep}{\epsilon}

\begin{document}
\begin{titlepage}

\begin{flushright}
\vbox{
\begin{tabular}{l}
UH-511-1123-08\\
 hep-ph/yymmnnn
\end{tabular}
}
\end{flushright}

\vspace{0.6cm}

\title{  ${\cal O}(\alpha_s^2)$ corrections to semileptonic 
decay $b \to c l \bar \nu_l$
}

\author{Kirill Melnikov \thanks{
e-mail:  kirill@phys.hawaii.edu}}
\affiliation{Department of Physics and Astronomy,\\ University of Hawaii,\\
Honolulu, HI, D96822}

\begin{abstract}

\vspace{2mm}

We compute the next-to-next-to-leading order (NNLO) QCD corrections to 
$b \to cl\bar \nu_l$ decay rate  at a fully differential level. Arbitrary 
cuts on kinematic variables of the decay products can be imposed.
Our  computation can be used  to study the  NNLO QCD corrections
to the total decay rate as well as to the lepton energy, 
hadronic invariant mass and hadronic energy moments 
and to incorporate those corrections into global fits 
of inclusive semileptonic $B$-decays.

\end{abstract}

\maketitle

\thispagestyle{empty}
\end{titlepage}

\section{Introduction}

Inclusive semileptonic decays of $B$-mesons into charmed final states 
are benchmark processes at $B$-factories. 
Because of  relatively large rates and 
clean experimental signatures, these decays can be studied
with great precision.  On the other hand, theoretical 
description of   semileptonic $B$ decays
is  robust thanks to the Operator Product Expansion (OPE) 
in inverse powers of the $b$-quark mass $m_b$. 
The application of the OPE to semileptonic decays of $B$-mesons
leads to the 
conclusion that both the total 
decay rate and various kinematic distributions can be described by 
power series in 
$\Lambda_{\rm QCD}/m_b$ \cite{ope}.
For infinitely heavy $b$-quark,  
the decay rate 
$B \to X_cl\bar \nu_l$ coincides with the  rate computed 
at the quark level. For realistic values of bottom and charm masses, 
a few non-perturbative matrix elements that enter at order 
$(\Lambda_{\rm QCD}/m_b)^n$, $n=2,3$ are accounted for in existing theoretical 
predictions.

In recent years, many  measurements of moments 
of charged lepton energy and hadronic invariant mass 
in  $B \to X_cl\bar \nu_l$ decays
have been performed
by BABAR, BELLE, CLEO, CDF and DELPHI
\cite{exp_babar,exp_del,exp_belle,exp_cleo,exp_cdf}.
Comparison of these  experimental results
with  theoretical predictions for corresponding  observables 
leads to the determination of   the Cabibbo-Kobayashi-Maskawa (CKM)
matrix element $|V_{\rm cb}|$,  
bottom and charm quark masses and a number of  
non-perturbative parameters such as $\mu_\pi^2$ 
and $\mu_G^2$ \cite{fit1,fit2}.  Typical precision claimed in these 
 analyses is about one percent for  $|V_{\rm cb}|$
and $m_b$ and a few percent for $m_c$ and non-perturbative 
matrix elements \cite{fit1,fit2}.

To achieve   such  precision,  advances 
in theoretical understanding  of semileptonic $B$-decays were necessary,
including 
subtle interplay between perturbative and non-perturbative physics and 
significant
developments in technology of multi-loop computations. While one-loop corrections to both $b \to cl\bar \nu_l$ total decay rate \cite{1lrate} and a number of important differential distributions are known since long  \cite{diffdistr}, it is interesting to remark 
that phenomenologically relevant triple  differential distribution in 
charged lepton energy, leptonic invariant mass and 
hadronic invariant mass was computed through ${\cal O}(\alpha_s)$ 
only a few  years ago \cite{trott,kolya1}.  This fact illustrates 
the complexity of perturbative calculations,  when  
 massive particles are involved, at a fully differential level.

Given the precision of available 
experimental measurements,  good understanding 
of non-perturbative effects and a fairly large value of the strong coupling 
constant $\alpha_s(m_b) \approx 0.24$, it is expected 
that ${\cal O}(\alpha_s^2)$ corrections to $b \to X_cl\bar \nu_l$ decays 
are required for a consistent theoretical description. However,
as was realized long ago, technical 
complexity of such an endeavor 
is daunting. To simplify the problem, the ${\cal O}(\alpha_s^2)$ 
corrections to $b \to X_cl \bar \nu_l$ 
were computed in three 
specific kinematic points \cite{mcz0,mczm,mczi}. These results were used 
in Ref.\cite{mczi} to estimate the  NNLO QCD corrections 
to $\Gamma(b \to X_cl\bar \nu_l)$.
Unfortunately,  such a description is necessarily limited in its scope even for the total rate and 
a generalization of such an approach to more differential quantities, such 
as lepton energy and hadronic invariant mass moments,  is  clearly out of question. 

On the other hand, a subset of the NNLO QCD corrections,  the  
BLM corrections \cite{blm}, received significant attention recently.
The BLM  corrections are associated with    the running 
of the strong coupling constant; they  are potentially important   since 
the QCD $\beta$-function is large.  
 For $B$ decays, however, the BLM corrections 
are known to be modest  if proper 
definition of quark masses is adopted and judicious choice of the 
renormalization scale in the strong coupling constant  is  made. 
The BLM effects are the easiest NNLO effects to calculate since they can 
be obtained from a one-loop computation if the latter is   performed 
with a non-vanishing gluon mass \cite{voloshin}. For this reason, 
in the past, the BLM 
corrections to $b \to cl\bar \nu_l$ were calculated  for  the total 
rate and  various kinematic moments \cite{wise,kolya2}. However, the 
NNLO QCD corrections  beyond the BLM 
approximation, for which genuine two-loop computations are required, remained 
missing.

Calculation  of these two-loop corrections  became possible recently
thanks to  developments in numerical approaches to multi-loop
computations \cite{method}.
These numerical methods benefit from the absence 
of mass hierarchy in the problem which is  the case for  
$b \to c$  decays, since masses of bottom and charm quarks are close.
The possibility to use the approach of Ref.\cite{method} to describe decays of 
charged particles  was recently pointed out in \cite{method1} where 
electron energy spectrum in muon decay was computed through 
second order in the perturbative expansion in QED.

The goal of this Letter is to present the computation of ${\cal O}(\alpha_s^2)$
corrections to $b \to X_cl\bar \nu_l$
 decay  rate at a fully differential level.
Our results can be used to calculate {\it arbitrary} observables related 
to inclusive $b \to c$ transition through NNLO in QCD.
For example, second order QCD corrections 
to such popular  observables as lepton energy,
hadronic invariant mass and hadronic energy moments can be studied 
in dependence 
of the  cut on charged lepton energy.  Inclusion of 
the results of our computation into global fits, should lead to a reduction 
of the theoretical uncertainty in the determination of $|V_{\rm cb}|$, 
the bottom and charm quark masses and the non-perturbative parameters that 
contribute to the decay rate.

\section{Computation}

In this Section, we set up our notation and briefly describe   
technical aspects of the computation. A detailed description of the 
method can be found in \cite{method,method1}.

Consider the decay $b \to X_cl\bar \nu_l$ where the final state lepton 
is massless. The  
differential decay rate can be written as 
\be
{\rm d} \Gamma = \frac{G_F |V_{\rm cb}|^2 m_b^5}{192 \pi^3}
\left ( {\rm d}F_0 + a\; {\rm d}F_1 + a^2\; {\rm d}F_2 \right ),
\ee
where $G_F$ is the Fermi constant, $m_b$ is the $b$-quark pole mass,
$a = \alpha_s/\pi$ 
and $\alpha_s$ is the ${\overline {\rm MS}}$ 
strong coupling constant defined in the theory with 
five active flavors and 
renormalized at the scale $m_b$. For numerical computations, we use $m_b = 4.6~{\rm GeV}$ and $m_c = 1.15~{\rm GeV}$. While these numerical values for 
the quark masses can not be justified in the pole scheme, our choice 
is motivated by an eventual necessity to transform the pole scheme computation
to a more suitable scheme. The values of the quark masses that we employ 
in this Letter correspond
to the central values of $m_{b,c}$ in the ``kinetic scheme'' \cite{kin}, 
derived in recent  fits to inclusive semileptonic $B$-decays \cite{fit1,fit2}.

To calculate the functions 
${\rm d}F_{0-2}$, we have to account for different  processes. 
At leading order, ${\rm d}F_0$ is computed by squaring  the matrix element of the 
process  $b \to cl\bar \nu_l$ and  summing  or averaging over spins and 
colors, as appropriate. At next-to-leading order, ${\rm d}F_1$ 
receives contributions 
from virtual ${\cal O}(\alpha_s)$ corrections to $b \to cl\bar \nu_l$
and from the real-emission process $b \to cl\bar \nu_l + g$. 
To compute ${\rm d}F_2$,  
we require two-loop ${\cal O}(\alpha_s^2)$ corrections 
to $b \to cl\bar \nu_l$, one-loop ${\cal O}(\alpha_s)$ corrections 
to $b \to cl\bar \nu_l + g$ and the  double real-emission corrections
$b \to cl\nu_l+X$, where  $X$ refers to  
two gluons or  a quark-antiquark pair or a ghost-antighost pair.
We will refer to these corrections as double-virtual, real-virtual and 
double-real, respectively. In addition, we have to account for a variety of 
renormalization constants, when computing higher order corrections.
We do not include the process $b \to cl \bar \nu_l + c \bar c$
in our calculation since the energy release in this process is so small 
that it can not be treated perturbatively.

To calculate the NNLO QCD corrections,
the method for multiloop computations 
developed in \cite{method,method1} is employed; in those references
a detailed discussion of many 
technical issues relevant for the current computation can be found.
One technical aspect that we improve upon  relative 
to Refs.\cite{method,method1} is how virtual corrections 
to single gluon emission process $b \to cl\bar \nu_l + g $  
are treated. In Refs. \cite{method,method1}
these corrections  
were dealt with by an analytic  reduction to master integrals followed 
by a numerical evaluation of those. This 
method, however, becomes impractical quite rapidly, once the number of 
external particles or the number of massive particles in the problem 
increases. In principle, 
the real-virtual corrections can be computed numerically, but for 
heavy-to-light decays this is  complicated  because 
some Feynman diagrams  develop imaginary parts. To handle these imaginary 
parts, we proceed as follows. For all Feynman 
diagrams that contribute to real-virtual corrections, 
it turns out possible to identify a Feynman parameter that enters 
the denominator of the integrand linearly. Let us call this Feynman parameter $x_1$. Then, 
a typical integral that has to be computed reads
\be
I(0,1) = \int \limits_{0}^{1} \frac{{\rm d}x_1 x_1^{-\ep + n}}{(-a + b x_1+i0)^{1+\ep}}.
\ee
Here $n \ge -1$, $b > a > 0$ and both $a$ and $b$ depend on other 
Feynman parameters 
and the kinematic variables. The two arguments of the 
function $I$ refer to  lower and upper limits  for $x_1$ integration.
To calculate $I(0,1)$, we
note that by extending upper integration boundary 
to infinity, a solvable
integral for arbitrary $a,b$ and $n$ is obtained. On the other hand, since
\be
I(0,1) = I(0,\infty) - I(1,\infty),
\ee
and because the denominator of the integrand in $I(1,\infty)$ is sign-definite,
$I(1,\infty)$ can be computed numerically in a straightforward way. It turns
out that, up to minor modifications, 
this trick  can be used to avoid dealing with 
the imaginary parts for all  Feynman diagrams that contribute 
to one-loop corrections to  single gluon emission process in $b \to c$ 
decays.

Because couplings of quarks and leptons  to the charged current are chiral, 
proper treatment of the Dirac matrix 
$\gamma_5$ in $d = 4-2\epsilon$ dimensions is important. While 
this problem  can be avoided in the computation of the 
total decay rate, for more differential 
quantities 
it becomes an issue.
We use the approach of Ref.\cite{larin} where a consistent framework 
for extending the axial vector current to $d$-dimensions is given. 

Our computation can be checked in a number of ways.
First, the double-virtual, 
real-virtual and double-real corrections are  divergent when taken separately but these divergences must  cancel in physical observables.
We have checked these cancellations for a variety of observables, 
from the  inclusive rate to various moments with cuts on both charged lepton 
energy and the hadronic invariant mass. Second, in the limit $m_c \to m_b$, 
the NNLO QCD corrections to the decay rate $b \to cl\bar \nu_l$ are described
by the so-called zero-recoil form factors computed through 
${\cal O}(\alpha_s^2)$ long time ago \cite{mcz0}. We have checked that in the 
limit $m_c \to m_b$ our computation reproduces the zero-recoil form factors.
Third, we can use published results  for 
the BLM corrections to the total rate and  charged lepton energy, hadronic 
invariant mass and hadronic energy moments \cite{kolya2} to check 
parts of our computation related to massless quark contributions to gluon 
vacuum polarization. Finally, considering the limit  $m_c \ll m_b$, we reproduce  the  NNLO QCD corrections to $b \to u l \bar \nu_l$
decay rate reported in Ref.\cite{timo}. 

\section{Results}

We are now in position to discuss  the results of our computation. We consider 
a number of observables, mostly for illustration purposes. 
We present the  results in the pole mass scheme and use the strong 
coupling constant renormalized at the scale $m_b$. While the pole mass 
scheme is known to be an unfortunate choice inasmuch as
the convergence of the perturbative expansion is concerned, 
we decided to present 
our results in this way for clarity. However, we emphasize that 
the   impact of the NNLO QCD  corrections, computed in this paper, 
on  the determination 
of $|V_{\rm cb}|$, heavy quark masses and the non-perturbative parameters, including kinetic and chromomagnetic heavy 
quark operators, can only be assessed once the
pole mass scheme is abandoned in 
favor of a more suitable quark mass definition and  the NNLO QCD corrections 
are included into the fit.

To present the results, we follow Ref.\cite{kolya2} and define
\ba
L_n (E_{\rm cut}) = \frac{
\langle (E_l/m_b)^n\;\theta(E_l -E_{\rm cut} )\; {\rm d}\Gamma \rangle }
{\langle {\rm d}\Gamma_0 \rangle },
\label{eq_3_1} \\
H_n (E_{\rm cut}) = \frac{
\langle (E_h/m_b)^n\;\theta(E_l -E_{\rm cut} )\; {\rm d}\Gamma \rangle }
{\langle {\rm d}\Gamma_0 \rangle },
\label{eq_3_1a}
\ea
where $\langle ... \rangle $ denotes average over the phase-space of all 
final state particles, 
$E_{l,h}$ is the energy of the charged lepton or hadronic system 
in the $b$-quark rest frame and 
\be
{\rm d} \Gamma_0 = \frac{G_F |V_{\rm cb}|^2 m_b^5}{192 \pi^3}\; {\rm d}F_0.
\label{eq_3_2}
\ee

The  lepton energy moments introduced in Eq.(\ref{eq_3_1}) can be written as
\be
L_n = L_n^{(0)} + a L_n^{(1)} + 
a^2 \left ( \beta_0 L_n^{(2,{\rm BLM})} + L_n^{(2)} \right )+...,
\ee
where ellipses stands for higher order terms in the perturbative expansion 
in QCD. Similar decomposition can be performed for  the 
hadronic energy moments $H_n$.
In addition, we use 
$\beta_0 = 11 - 2/3 N_f$ and define the non-BLM corrections $L_n^{(2)},H_n^{(2)}$ as 
the difference of the complete ${\cal O}(\alpha_s^2)$ correction 
and the BLM correction computed with   $N_f = 3$.  

\begin{tiny}
\begin{table}[htbp]
\vspace{0.1cm}
\begin{center}
\begin{tabular}{|c|c|c|c|c|c|}
\hline\hline
$n$ & $E_{\rm cut}$, GeV & $L_n^{(0)}$ & $L_n^{(1)}$ & $L_n^{(2,{\rm BLM})}$ & $L_n^{(2)}$ \\ \hline\hline
$0$ & $0$ &  $1$ & -1.77759 & $-1.9170$ & $3.40$ \\ \hline
$1$ & $0$ & $0.307202$ & $-0.55126 $ & $-0.6179$ & $1.11 $ \\ \hline 
$2$ & $0$ & $0.10299$ & $-0.1877 $ & $-0.2175$ & $0.394 $ \\ \hline \hline
$0$ & $1$& $0.81483$ & -1.4394 & $-1.5999$ & $2.63$ \\ \hline
$1$ & $1$& $0.27763$ & -0.49755 & $-0.5667$ & $1.00$ \\ \hline
$2$ & $1$& $0.09793$ & -0.17846 & $-0.20875$ & $0.382$ \\ \hline \hline 
\end{tabular}
\caption{\label{table1}  Lepton energy moments.}
\vspace{-0.1cm}
\end{center}
\end{table}
\end{tiny}

In Tables~\ref{table1},\ref{table2} 
the results for  lepton energy and hadronic energy 
moments with and without a 
cut  on the lepton energy are displayed.  The numerical accuracy of 
$L_n^{(0,1)},H_n^{(0,1)}$ and $L_n^{(2,\rm BLM)},H_n^{(2,\rm BLM)}$ is about $0.1-0.2\%$ whereas
the numerical accuracy of $L_n^{(2)},H_n^{(2)}$ is about $1-3\%$. It is possible to
improve on the accuracy but this requires somewhat large CPU time. Nevertheless,
for all practical applications the achieved numerical accuracy is sufficient.

 There are a few interesting observations 
that follow from Tables~\ref{table1},\ref{table2}.
Quite generally, the non-BLM corrections and the BLM corrections 
have opposite signs; given their relative magnitude and 
the value of $\beta_0$,  it is easy to see that  the ${\cal O}(\alpha_s^2)$
corrections 
are about twenty percent  smaller than what the BLM-based estimates 
suggest.   The relative magnitude of the non-BLM and BLM corrections
is largely independent of $n$ and of whether   the 
lepton energy cut is applied.

\begin{tiny}
\begin{table}[htbp]
\vspace{0.1cm}
\begin{center}
\begin{tabular}{|c|c|c|c|c|c|}
\hline\hline
$n$ & $E_{\rm cut}$, GeV & $H_n^{(0)}$ & $H_n^{(1)}$ & $H_n^{(2,{\rm BLM})}$ & $H_n^{(2)}$ \\ \hline\hline
$1$ & $1$& $0.334$ & -0.57728 & $-0.6118$ & $1.02$ \\ \hline
$2$ & $1$& $0.14111$ & -0.23456 & $-0.2343$ & $0.362$\\ \hline \hline
\end{tabular}
\caption{\label{table2}  Hadronic energy moments.}
\vspace{-0.1cm}
\end{center}
\end{table}
\end{tiny}

First row  in Table~\ref{table1} provides the NNLO QCD corrections to 
the total decay rate $b \to cl\bar \nu_l$ in the pole  mass scheme.
Such corrections were estimated earlier in Ref.\cite{mczi}. Note that 
in Ref.\cite{mczi}  the numerical  results are given for the 
ratio of quark masses $m_c/m_b = 0.3$ and also 
the BLM corrections are  defined with $N_f=4$, rather than $N_f = 3$. 
Calculating the non-BLM corrections for the set of parameters
employed in \cite{mczi}, we find 
$L_0^{(2)} \approx 1.73$ which is to be compared with the estimate 
$L_0^{(2)} \approx 0.9(3)$, reported in \cite{mczi,comm}.

The results of Ref.\cite{mczi} were used in Ref.\cite{kolya3} to estimate the 
impact of the QCD corrections on $\Gamma(B \to X_c l \nu_l)$. 
In Ref.\cite{kolya3} the perturbative corrections to $b \to c l \bar \nu_l$
decay rate
are described  by a factor $A^{\rm pert}$, defined as 
\be
\Gamma(b \to X_c l \bar \nu_l) = A^{\rm pert}(r) \; \langle  {\rm d} \Gamma_0 \rangle,
\ee
where $r = m_c/m_b$. 
$A^{\rm pert}$ depends on the adopted scheme for the quark masses. In the 
kinetic mass scheme, $A^{\rm pert}(0.25) = 0.908$ is quoted.
To arrive at this result,
Ref.\cite{kolya3} uses $L_0^{(2)} =1.4$ which 
is about a factor $2.5$ smaller than the corresponding entry 
in Table~\ref{table1}.  Correcting for this discrepancy, we derive 
\be
A^{\rm pert}(0.25) = 0.919.
\ee
We believe that this  value for the perturbative renormalization factor in the kinetic scheme for $m_c/m_b = 0.25$ should be employed 
in global fits of semileptonic $B$-decays.

Further analysis of entries in Table~\ref{table1} suggests that 
the QCD corrections in general and the non-BLM corrections in 
particular mostly affect the  overall normalization
 rather than shapes of kinematic distributions. This follows from 
the approximate independence  of $L_n^{(1,2)}/L_n^{(0)}$ of $n$ and 
also of whether or not the cut on the lepton energy is imposed.
It is therefore possible to speculate that 
the non-BLM corrections computed in this Letter will mostly affect
the extraction of $|V_{\rm cb}|$ whereas their influence on, e.g.,  
the $b$-quark mass determination  
will be  minor. Concerning the $|V_{\rm cb}|$, the 
increase of the perturbative renormalization factor $A_{\rm pert}$ by 
$10 \times 10^{-3}$ implies the change  in the value of 
$|V_{\rm cb}|$, extracted in Ref.\cite{fit2}, by about 
$-0.25 \times 10^{-3}$. On the other hand, since non-BLM corrections 
were not included in a  fit of Ref.\cite{fit1}, the shift in the 
value of $|V_{\rm cb}|$ derived in that reference will likely 
be larger $\sim -0.5 \times 10^{-3}$. Although  expected 
 shifts in central values 
of $|V_{\rm cb}|$ are not large, we stress that they are comparable 
to uncertainties in $|V_{\rm cb}|$, derived  in the 
global fits \cite{fit1,fit2}.

\section{Conclusions}

In this Letter, the computation of the NNLO QCD corrections to the fully differential $b \to cl\bar \nu_l$ decay rate  is reported. 
The differential nature of the 
computation makes it possible to apply arbitrary cuts on the kinematic
variables of final state particles.  This result allows to extend the existing 
determinations of the CKM matrix element $|V_{\rm cb}|$, the bottom and charm 
quark masses and the non-perturbative parameters $\mu_\pi^2$ and $\mu_G^2$
from global fits to semileptonic decays of $B$-mesons, by including the 
NNLO QCD corrections {\it exactly}. We note that for a consistent 
high-precision analysis of semileptonic $B$-decays, 
also ${\cal O}(\alpha_s)$  corrections 
to Wilson coefficients  of non-perturbative kinetic and chromomagnetic 
operators are required. Such a correction is available for the kinetic 
operator \cite{tb} but is still missing for the chromomagnetic.

We presented a few results for charged lepton energy moments and hadronic 
energy moments with and without a cut on the lepton energy in the 
pole mass scheme. These results suggest 
that 
the magnitude of the non-BLM corrections does not depend strongly on the 
kinematics; the non-BLM corrections 
are approximately $2\%$ for all the moments considered. 
We therefore expect that the non-BLM NNLO QCD corrections will mostly 
affect the determination of $|V_{\rm cb}|$ decreasing its central value 
by about one percent whereas their impact on the quark masses 
and the non-perturbative parameters will  probably be 
quite mild.  

As a final remark, 
we note that it would be interesting to extend this calculation 
in two ways. First, one may consider semileptonic decays of 
$B$-mesons into massive leptons. Such an extension, relevant 
for the description of $B \to X_c + \tau + \bar \nu_\tau$ decay 
is straightforward. Second,  it is interesting to extend the 
current calculations to allow for a {\it massless} quark in the 
final state. This is  a  difficult problem but it is highly 
relevant for the determination of the CKM matrix element 
$|V_{\rm ub}|$ from semileptonic $b \to u$ transitions.

\vspace*{0.2cm}
{\bf Acknowledgments} 
Discussions with F.~Petriello and useful correspondence with T.~Becher
are gratefully acknowledged. I would like to thank A.~Czarnecki and 
A.~Pak for informing me about their results prior to publication.
This research is partially supported 
by the DOE under grant number 
DE-FG03-94ER-40833.


\begin{thebibliography}{99}

\bibitem{ope} M.~A.~Shifman and M.~B.~Voloshin, Sov. J. Nucl. Phys. {\bf 41}, 120 (1985); J.~Chay, H.~Georgi and B.~Grinstein, Phys. Lett. {\bf B247}, 
399 (1990); I.~I.~Bigi, N.~Uraltsev and A.~Vainshtein, Phys. Lett. {\bf B293}, 430 (1992) [Erratum: {\bf B297},  477 (1992)]; I.I.~Bigi, M.~Shifman, N.~Uraltsev and A.~Vainshtein, Phys. Rev. Lett. {\bf 71}, 496 (1993).

\bibitem{exp_babar} B.~Aubert {\it et al.} (BABAR Collaboration), Phys. Rev. Lett. {\bf 93}, 011803 (2004); Phys. Rev. {\bf D69}, 111103 (2004); 
Phys. Rev. {\bf D69}, 111104 (2004);
Phys. Rev. {\bf D72}, 052004 (2005); hep-ex/0507001.

\bibitem{exp_del}
J.~Abdallah {\it et al.} (DELPHI Collaboraition), 
Eur. Phys. J. {\bf C45}, 35 (2006); 

\bibitem{exp_belle} K.~Abe {\it et al.} (Belle Collaboration), Phys. Rev. Lett.
{\bf 93}, 061803 (2004); hep-ex/0508005.

\bibitem{exp_cleo} S.~E.~Csorna {\it et al.} (CLEO Collaboration), Phys. Rev. {\bf D70}, 032002 (2004); S.~Chen {\it et al.} (CLEO Collaboration), Phys. Rev. Lett. {\bf 87}, 251807 (2001).

\bibitem{exp_cdf} D.~Acosta {\it et al.} (CDF Collaboration), Phys. Rev. {\bf D71}, 051103 (2005).

\bibitem{fit1} C.W.~Bauer, Z.~Ligeti, M.~Luke and A.V.~Manohar, 
Phys. Rev. {\bf D67}, 054012 (2003).

\bibitem{fit2}  O.I.~Buchm\"uller and H.~U.~Fl\"acher, 
Phys. Rev. {\bf D73}, 073008 (2006).

\bibitem{1lrate} Y.~Nir, Phys. Lett. {\bf B221}, 184 (1989).

\bibitem{diffdistr} A.~Ali and E.~Pietranen, Nucl. Phys. {\bf B154}, 519 (1979);
G.~Altarelli, N.~Cabbibo, G.~Corbo, L.~Maiani and G.~Martinelli, Nucl. Phys. {\bf B 208}, 365 (1982); M.~Jezabek and J.~H.~K\"uhn, Nucl. Phys. {\bf B314}, 1 (1989);
Nucl. Phys. {\bf B320}, 20 (1989); A.~Czarnecki, M.~Jezabek and J.~H.~K\"uhn, 
Acta Phys. Polonica {\bf B20}, 961 (1989); A.~Czarnecki and M.~Jezabek, Nucl. Phys. {\bf B427}, 3 (1994); A.~Falk, M.~E.~Luke and M.~J.~Savage, Phys. Rev. {\bf D53}, 2491 (1996);
C.~W.~Bauer and B.~Grinstein, Phys. Rev. {\bf D68} 054002 (2003);
M.~B.~Voloshin, Phys. Rev. {\bf D51} 4934 (1995); A.~F.~Falk and M.~E.~Luke 
Phys. Rev. {\bf D57}, 424 (1998).

\bibitem{trott} M.~Trott, Phys. Rev. {\bf D70}, 073003 (2004).
\bibitem{kolya1} N.~Uraltsev, Int. J. Mod. Phys. {\bf A20}, 2099 (2005). 

\bibitem{mczi} A.~Czarnecki and K.~Melnikov, Phys. Rev. {\bf D59}, 014036 (1999).

\bibitem{comm}
The value of $L_0^{(2)}$ quoted 
in Eq.(10) of Ref.~\cite{mczi} is $1.4$ but this value is in error 
due to an unfortunate mistake in the interpolating procedure.  
Correcting for
this mistake, one obtains $L_0^{(2)} = 0.9(3)$.

\bibitem{mczm} A.~Czarnecki and K.~Melnikov, Phys. Rev. Lett. {\bf 78}, 3630 (1997).

\bibitem{mcz0} A.~Czarnecki, Phys. Rev. Lett. {\bf 76}, 4124 (1996);
A.~Czarnecki and K.~Melnikov, Nucl. Phys. {\bf B505}, 65 (1997);
J.~Franzkowski and J.~B.~Tausk, Eur. Phys. J. {\bf C5}, 517 (1998).

\bibitem{blm} S.~J.~Brodsky, G.~P.~Lepage and P.~B.~Mackenzie, 
Phys. Rev. {\bf D28}, 228 (1983).

\bibitem{voloshin} B.~H.~Smith and M.~B.~Voloshin, Phys. Lett. {\bf B340}, 176 (1994).

\bibitem{wise} M.~Luke, M.~J.~Savage and M.~B.~Wise, 
Phys. Lett. {\bf B345}, 301 (1995).

\bibitem{kolya2} V.~Aquila, P.~Gambino, G.~Ridolfi and N.~Uraltsev, 
Nucl. Phys. {\bf B719}, 77 (2005).

\bibitem{method} 
C.~Anastasiou, K.~Melnikov and F.~Petriello, Phys. Rev. {\bf D69}, 076010 
(2004); Phys. Rev. Lett. {\bf 93}, 032002 (2004);
Phys. Rev. Lett. {\bf 93}, 262002 (2004);
Nucl. Phys. {\bf B724}, 197 (2005).

\bibitem{method1}
C.~Anastasiou, K.~Melnikov and F.~Petriello, 
  JHEP {\bf 0709}, 014 (2007).



\bibitem{kin} I.~Bigi, M.~Shifman, N.~Uraltsev and A.~Vainshtein, 
Phys. Rev. {\bf D56}, 4017 (1997); N.G.~Uraltsev, Nucl. Phys. {\bf B491}, 303
(1997).

\bibitem{larin}
S.~Larin, Phys. Lett. {\bf B303}, 113 (1993).


\bibitem{timo} T. van Ritbergen, Phys. Lett. {\bf B454}, 353 (1999). 

\bibitem{kolya3} D.~Benson, I.~I.~Bigi, Th.~Mannel and N.~Uraltsev, 
hep-ph/0302262.

\bibitem{tb} T.~Becher, H.~Boos and E.~Lunghi, 
JHEP {\bf 0712}, 062 (2007).

\end{thebibliography}
\end{document}